\documentclass[conference]{IEEEtran}
\usepackage{cite}
\usepackage{amsmath,amssymb,amsfonts}
\usepackage{algorithmic}
\usepackage[linesnumbered,ruled,vlined]{algorithm2e}
\usepackage{graphicx}
\usepackage{textcomp}
\usepackage{xcolor}

\usepackage[flushleft]{threeparttable}
\usepackage{array,booktabs,makecell}
\usepackage{comment}
\usepackage{subcaption}
\usepackage{amsthm}
\usepackage{nicefrac}
\allowdisplaybreaks

\def\BibTeX{{\rm B\kern-.05em{\sc i\kern-.025em b}\kern-.08em
    T\kern-.1667em\lower.7ex\hbox{E}\kern-.125emX}}
\begin{document}

\title{Priority and Stackelberg Game-Based Incentive Task Allocation for Device-Assisted MEC Networks}

\author{\IEEEauthorblockN{Yang Li\IEEEauthorrefmark{2}, Xing Zhang\IEEEauthorrefmark{2}\,\IEEEauthorrefmark{1}, Bo Lei\IEEEauthorrefmark{3}, Zheyan Qu\IEEEauthorrefmark{2}, Wenbo Wang\IEEEauthorrefmark{2} }
\IEEEauthorrefmark{2}Key Laboratory of Universal Wireless Communications, Ministry of Education\\
\IEEEauthorrefmark{2}Beijing University of Posts and Telecommunications, Beijing 100876, China \\
\IEEEauthorrefmark{3}Beijing Branch of China Telecom Co., Ltd., Beijing 100032, China \\
\IEEEauthorrefmark{1}Email: zhangx@ieee.org
}

\maketitle

\begin{abstract}
Mobile edge computing (MEC) is a promising computing paradigm that offers users proximity and instant computing services for various applications, and it has become an essential component of the Internet of Things (IoT). However, as compute-intensive services continue to emerge and the number of IoT devices explodes, MEC servers are confronted with resource limitations. In this work, we investigate a task-offloading framework for device-assisted edge computing, which allows MEC servers to assign certain tasks to auxiliary IoT devices (ADs) for processing. To facilitate efficient collaboration among task IoT devices (TDs), the MEC server, and ADs, we propose an incentive-driven pricing and task allocation scheme. Initially, the MEC server employs the Vickrey auction mechanism to recruit ADs. Subsequently, based on the Stackelberg game, we analyze the interactions between TDs and the MEC server. Finally, we establish the optimal service pricing and task allocation strategy, guided by the Stackelberg model and priority settings. Simulation results show that the proposed scheme dramatically improves the utility of the MEC server while safeguarding the interests of TDs and ADs, achieving a triple-win scenario.
\end{abstract}


\section{Introduction}
With the rapid development of the Internet of Things (IoT) and 5G technology, there is an increasing need for computation-intensity and time-sensitive tasks. To address the need for real-time communication and computation in various emerging services, mobile edge computing (MEC) has emerged as a promising paradigm within IoT. It involves deploying computation, storage, and network services near IoT devices to support real-time processing of tasks. However, the increasing quantity of IoT devices and the demand for task processing could overwhelm MEC installations. Meanwhile, cost constraints limit the expansion of computation and storage capacities of edge servers (ESs). Additionally, IoT devices' continuously improving computation capabilities and storage capacities represent valuable resources that can enhance MEC \cite{2}. Device-assisted MEC, as a subset of MEC, is garnering increasing attention from researchers, and much work has been carried out to investigate the issues in device-assisted MEC \cite{3,17,16}.

% Regarding latency reduction, the authors in \cite{3} considered a Device-to-Device (D2D)-enabled MEC offloading scenario, where a device can partially offload its computation task to the ES or exploit the computation resources of proximal devices. They jointly optimized partial offloading and resource allocation to reduce task execution delay. Feng \textit{et al.} \cite{4} introduced a reverse offloading framework designed to fully harness the computational resources of vehicles, thereby alleviating the strain on the vehicular edge computing server and reducing system latency. For energy consumption reduction, the work in \cite{6} proposed a low-complexity heuristic resource allocation strategy to tackle the challenge of energy-efficient resource allocation in a multi-device D2D-assisted fog computing scenario. Concerning capacity enhancement, the study \cite{7} focused on maximizing the overall computing capacity of a device-enhanced MEC system. The optimization problem aimed to maximize the number of supported devices in the system, considering the constraints of communication and computational resources. The authors decomposed this problem into two sub-problems for easier resolution.

However, none of the aforementioned works have addressed incentive design. Indeed, task IoT devices (TDs), auxiliary IoT devices (ADs), and ESs represent typical profit-driven entities with conflicting interests. Effective collaboration among the three parties may be impeded in the absence of a proper incentive mechanism. Therefore, some research has focused on motivating the diverse participants in MEC by developing incentive mechanisms. The authors in \cite{14,15} explored the optimization of resource allocation via pricing mechanisms in MEC systems. The study in \cite{10} devised a blockchain-based resource transaction framework, which utilizes the blockchain's features of decentralization, immutability, and smart contracts for secure resource transactions. To effectively incentivize user devices to act as computation providers (CPs) for computation requestors (CRs), Chen \textit{et al.} \cite{11} introduced an incentive mechanism based on contract theory to mitigate the information asymmetry issue in D2D computation offloading. However, the above studies exclusively focus on interactions between the ES and TDs or the ES and ADs, neglecting the scenario of interactions among all three parties. Consequently, none of the above incentive mechanisms can be effectively employed in device-assisted MEC scenarios.

In response to the limitations of current research, we have developed an incentive-driven pricing and task allocation algorithm tailored to device-assisted MEC scenarios. Specifically, we initially utilize the Vickrey auction mechanism to incentivize ADs. Subsequently, we utilize the Stackelberg game to analyze the interaction process between TDs and the ES. Lastly, building upon the Stackelberg model and priority settings, we introduce a service pricing and task allocation algorithm aimed at maximizing the utility of the ES. In addition, we conduct extensive simulations under different parameter settings to demonstrate the effectiveness of our incentive mechanism.

\section{System Models}\label{sec:models}
\begin{figure}[htbp]
    \centerline{\includegraphics[width=0.9\linewidth]{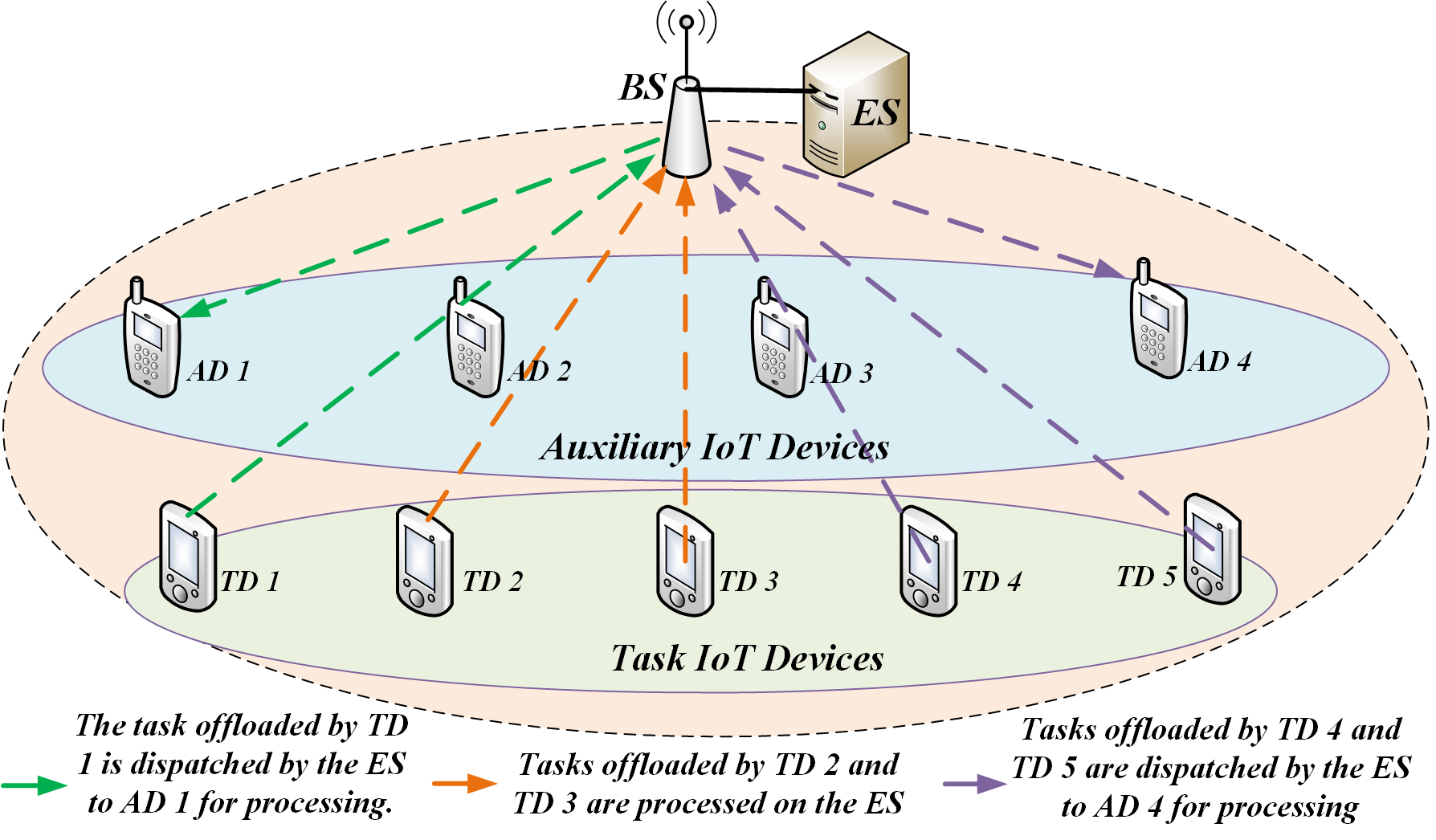}}
    \vspace{-0.2cm}
    \caption{An illustration of the device-assisted MEC network.}
    \label{fig:system}
    \vspace{-0.1cm}
\end{figure}

\vspace{-0.15cm}
This section presents the model of a device-assisted MEC network, as depicted in Fig. $\ref{fig:system}$. The scenario involves the edge server (ES), task IoT devices (TDs), and auxiliary IoT devices (ADs), which are described as follows.

\textit{(1) Edge server:} An ES is linked with a base station (BS) via a wired connection, offering edge computing services to users within the BS-managed cell. When the ES is overloaded, it can recruit some ADs to handle the overloaded tasks, such as TD $1$, TD $4$, and TD $5$ in Fig. $\ref{fig:system}$, but it should pay rewards to the ADs that handle the tasks.

\textit{(2) Task IoT device:} TDs are devices that request task offloading from the ES. Assuming TDs act rationally, they offload some or all data to the ES based on its pricing to maximize their own utility.

\textit{(3) Auxiliary IoT devices:} During recruitment initiated by the ES, ADs can provide information such as their location and available resources, as well as submit bids at the lowest acceptable price per CPU cycle. This allows them to benefit by assisting the ES in handling tasks.
\vspace{-0.45cm}

The system is assumed to have $N$ TDs and $M$ ADs, denoted by the sets $\mathcal{N}$ and $\mathcal{M}$, respectively. The task generated by TD $i\in \mathcal{N}$ is denoted as $H_i \triangleq \{L_i, C_i, t_i^{max}\}$, where $L _i$ represents the size of the task $H_i$, $C_i$ is the required number of CPU cycles to complete $H_i$, and $t_i^{max}$ indicates the time constraint of $H_i$. Similar to \cite{14}, we introduce a complexity factor $\phi_i$ to quantify the CPU cycles needed to process one bit of $H_i$. The offloaded portion of TD $i$'s task is denoted as $h_i\triangleq \{l_i, c_i, t_i^{max}\}$, where $l_i$ signifies the size of $h_i$, $c_i$ represents the required number of CPU cycles to complete $h_i$, and $t_i^{max}$ indicates the time constraint of $h_i$. Based on the complexity factor $\phi_i$, we can obtain $c_i = \phi_il_i$, where $l_i \in [0,L_i]$ and $c_i \in [0,C_i]$.

\subsection{Communication Model}\label{sec:comm_model}

We use the same communication model as in \cite{13}, and the transmission delay between the BS and ES is disregarded. The transmission rate from TD $i$ to the ES is defined as $R_i^B = W_B log_2(1+\frac{p_ig_i^B}{N_0})$, where $W_B$ is the channel bandwidth between TD $i$ and the BS. $p_i$ is the transmit power of TD $i$, and $N_0$ is the background noise power. $g_i^B = \mu_0(d_i^B)^{\tau}$ is the channel gain between TD $i$ and the BS, where $\mu_0$ is the fading component, $\tau$ is the channel path loss exponent, and $d_i^B$ is the distance between TD $i$ and the BS.

Additionally, when the ES schedules the offloaded task $h_i$ for processing on AD $j$, the transmission rate from the BS to AD $j$ is defined as $R_B^j = W_B log_2(1+\frac{p_Bg_j^B}{N_0})$, where $p_B$ is the transmit power of the BS. $g_i^B = \mu_0(d_j^B)^{\tau}$ is the channel gain between the BS and AD $j$, and $d_j^B$ is the distance between AD $j$ and the BS.

\subsection{Computation Model}\label{sec:comp_model}
The ES determines whether each offloaded task is computed locally or scheduled for processing on an AD. For offloaded task $h_i$, we use $x_i\in\{0,1,2, ..., M\}$ to indicate its processing location. $x_i = 0$ indicates that task $h_i$ is processed on the ES, while $x_i = m \in \{1,2,... ,M\}$ indicates that task $h_i$ is processed on AD $m$. Subsequently, we elaborate on each of these two scenarios.

\textit{(1) Processing on the ES:} Based on the task model and communication model, the transmission time of task $h_i$ from TD $i$ to the ES can be calculated as $t_{i,B}^{tran}=\frac{\boldsymbol{1}_{x_i=0}l_i}{R_i^B}$, where $\boldsymbol{1}_{\{\cdot\}}$ is the indicator function and equals 1 (resp., 0) if the condition is true (resp., false).

Given that the time constraint of task $h_i$ is $t_i^{max}$, the processing time of $h_i$ on the ES must not exceed $t_i^{max}-t_{i, B}^{tran}$. Therefore, the minimum computational resources allocated by the ES for TD $i$ can be calculated as 
\begin{align}
f_i^B=\frac{\boldsymbol{1}_{x_i =0}c_i}{t_i^{max}-t_{i, B}^{tran}}.
\label{eq:f_i^B}
\end{align}

\textit{(2) Processing on AD $j$:} Similarly, the transmission time of task $h_i$ to AD $j$ through the BS relay can be calculated as $t_{i,j}^{tran}= \frac{l_i}{R_i^B}+\frac{l_i}{R_{B}^j}$. Therefore, the minimum computational resources allocated by AD $j$ for TD $i$ can be calculated as 
\vspace{-0.1cm}
\begin{align}
f_i^j=\frac{\boldsymbol{1}_{{x_i=j}}c_i}{t_i^{max}-t_{i,j}^{tran}}.
\label{eq:f_i^j}
\end{align}

In this study, we assume that the ES and ADs handle task $h_i$ with the minimal computational resources necessary to meet its time constraint, motivated by their respective interests.

\subsection{Utility Model}\label{sec:util_model}
\textit{(1) TD $i$’s model:} For TD $i$, we employ a logarithmic utility function to represent the satisfaction derived from offloading computation tasks, which can properly reflect the relationship between TD $i$ and satisfaction \cite{14}. This function can be expressed as
\vspace{-0.3cm}
\begin{align}
s_i = w_i ln(1+l_i),
\label{eq:s_i}
\vspace{-0.3cm}
\end{align}
where $w_i$ denotes the satisfaction factor of TD $i$. Clearly, as TD $i$ offloads more data, both its own battery life and hardware wear decrease, thereby leading to a greater value of the satisfaction function.

Naturally, when a TD offloads its computational task to the ES, it incurs a fee. Let's assume that for TD $i$, the ES imposes a fixed fee of $d_i$ per CPU cycle required to process its task. Consequently, the payment from TD $i$ to the ES can be calculated as
\vspace{-0.3cm}
\begin{align}
o_i = d_i\phi_il_i.
\label{eq:o_i}
\end{align}

Additionally, TD $i$ incurs energy consumption for both local processing and offloading tasks, with its energy cost calculated as
\vspace{-0.3cm}
\begin{align}
e_i = \gamma q_i \phi_i(L_i-l_i)+\gamma p_i\frac{l_i}{R_i^B},
\label{eq:e_i}
\vspace{-0.3cm}
\end{align}
where $q_i$ denotes the energy consumption per CPU cycle when TD $i$ processes a task, and $\gamma$ denotes the cost per unit of energy consumption.

Thus, the utility of TD $i$ can be expressed as
\begin{align}
U_i = s_i+v_i-e_i-o_i,
\label{eq:U_i}
\end{align}
where $v_i$ represents the value to the user when the task is successfully processed.

\textit{(2) ES's model:} For the offloaded task $h_i$, if the ES assigns the task to be processed locally, the energy cost is calculated as $e_B^i = \gamma q_B\phi_il_i$, where $q_B$ represents the energy consumption per CPU cycle when the ES processes a task\footnote{Since devices need to find a balance between power consumption and performance, they generally use processors with lower power consumption and relatively lower performance. In contrast, ESs, which frequently handle computationally intensive tasks and have access to a constant power supply, tend to utilize higher-performance processors. Therefore, $q_B>q_i, \forall i \in \mathcal{N}$.}.

Thus, the utility of the ES for handling task $h_i$ can be expressed as
\vspace{-0.3cm}
\begin{align}
U_{B,l}^i = d_i\phi_il_i - \gamma q_B\phi_il_i.
\label{eq:U_{B,l}^i}
\end{align}

When the ES assigns task $h_i$ to AD $j$ for processing, its utility is calculated as the revenue generated minus the payment made to AD $j$ and the energy cost incurred for forwarding that task. This can be expressed as
\begin{align}
U_{B,j}^i = d_i\phi_il_i - d_j\phi_il_i-\gamma \frac{p_Bl_i}{R_B^j},
\label{eq:U_{B,j}^i}
\end{align}
where $d_j$ represents the reward obtained per CPU cycle when AD $j$ assists ES in processing a task.

In summary, the total utility of the ES can be expressed as
\begin{align}
U_B = \sum_{i=1}^N (\boldsymbol{1}_{x_i=0}U_{B,l}^i+\sum_{j=1}^M \boldsymbol{1}_{x_i=j} U_{B,j}^i).
\label{eq:U_B}
\end{align}

\textit{(3) AD $j$'s model:} Assuming that AD $j$ requires $C$ CPU cycles to process tasks allocated by  the ES, and its bid is denoted as $a_j$, its utility can be expressed as $U_j = (d_j-a_j)C$.

\section{Problem Formulation and Analysis}\label{sec:problem}
\subsection{Pricing for ADs}
The recruitment of ADs for the ES can be conceptualized as an auction game. Among ADs meeting the resource requirements of task $h_i$, the one which brings the highest utility to the ES wins the auction, serves as the processing device for the task, and earns a reward. Implementing the Vickrey auction can incentivize ADs to bid honestly \cite{15}. In this auction format, the pricing strategy employed by the ES for AD $j$ involves setting the price at the lowest bid among those exceeding AD $j$'s offer. At this point, the AD's individual rationality is satisfied, i.e., its utility is non-negative.

The interactions between the ES and TDs can be modeled as a Stackelberg game. The subsequent subsection introduces the optimization problem based on the Stackelberg game and offers an analysis.

\begin{figure}[htbp]
    \vspace{-0.5cm}
    \centerline{\includegraphics[width=0.8\linewidth]{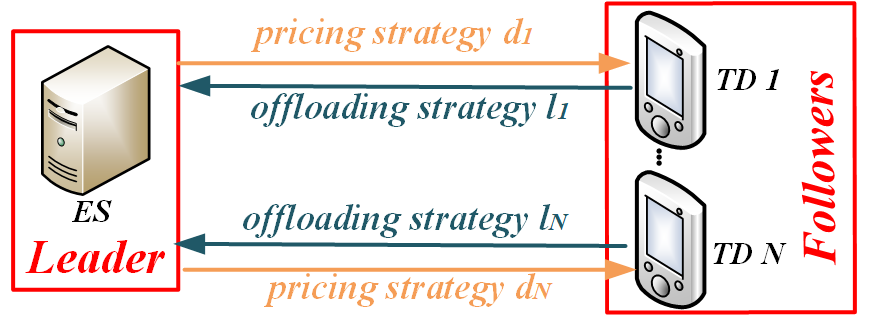}}
    \vspace{-0.0cm}
    \caption{Stackelberg game model.}
    \label{fig:game model}
    \vspace{-0.5cm}
\end{figure}
\subsection{Problem Formulation}\label{sec:formulation}
Fig. $\ref{fig:game model}$ illustrates the Stackelberg game involving the ES and TDs. In this game, the ES, acting as the leader, sets its pricing policy $d_i$ for TD $i$, $\forall i \in \cal N$, while TD $i$, as a follower, decides the size of task $h_i$ to offload. The pricing strategy and offloading strategy are denoted as $\boldsymbol{d}=(d_1,d_2,.... ,d_N)$ and $\boldsymbol{l}=(l_1,l_2,.... ,l_N)$, respectively. Since the pricing strategy is formulated by the ES, the goal of the pricing strategy is to maximize the utility of the ES. However, in the scenario we consider, the ES must not only determine the pricing strategy but also assign processing locations for all offloaded tasks. Therefore, the problem of maximizing the utility of the ES can be formulated as 
\vspace{-0.2cm}
\begin{align}
\mathcal{P}_1: \quad &\max_{\boldsymbol{d,x}} U_B \\
\text{s.t.} \quad &(10a):d_i^{min} \le d_i \le d_i^{max}, \forall i \notag \\
&(10b):x_i \in \{0,1,...,M\}, \forall i \notag \\
&(10c):\sum_{i=1}^N f_i^B\boldsymbol{1}_{x_i=0} \le F_B \notag \\
&(10d):\sum_{i=1}^N f_i^j \boldsymbol{1}_{x_i=j} \le F_j, \forall j \notag \\
&(10e):\boldsymbol{1}_{x_i=0}U_{B,l}^i + \sum_{j=1}^M \boldsymbol{1}_{x_i=j}U_{B,j}^i\ge 0, \forall i,\notag 
\end{align}
where $\boldsymbol{x}=\{x_1,x_2,.... .x_N\}$ denotes the task processing locations assigned by the ES for all TDs. $F_B$ and $F_j$ denote the overall idle resources of the ES and AD $j$, respectively. Constraint (10a) shows that the pricing strategy of the ES is within a specified range, where $d_i^{max}$ and $d_i^{min}$ represent the maximum and minimum pricing strategies for TD $i$, respectively. Constraint (10b) states that each offloaded task is processed at the ES or an AD. Constraints (10c) and (10d) specify that the total computational resources allocated by the ES and ADs cannot exceed their idle resources. Constraint (10e) specifies that the service pricing and task allocation must satisfy the individual rationality of the ES.

Subsequently, TD $i$ determines the amount of offloaded data $l_i$ to maximize its utility after being informed of the ES's pricing strategy. Thus, the optimization problem for TD $i$ is expressed as
\vspace{-0.4cm}
\begin{align}
\mathcal{P}_2: \quad &max_{{l_i}} U_i \\
\text{s.t.} \quad &(11a):0\le l_i\le L_i \notag \\
&(11b):U_i \ge 0, \notag
\end{align}
where (11a) constrains the amount of offloaded data $l_i$, while constraint (11b) states that TD $i$ is a rational entity.

\subsection{Analysis of Game Problem}\label{sec:analysis}
In this section, we initially assume that the ES possesses ample computational resources, meaning it doesn't delegate tasks to ADs due to overload. Subsequently, upon attaining the optimal solution based on this assumption, we account for the resource constraints of the ES and refine the solution accordingly.

Given the current assumption, we can utilize the backward induction method to analyze the proposed problem. In the first stage, the optimal offloading decision for each TD is determined. In the second stage, according to the optimal strategy of all TDs, we ascertain the optimal pricing strategy for the ES.

Initially, by referring to equations ($\ref{eq:s_i}$), ($\ref{eq:o_i}$), ($\ref{eq:e_i}$), and ($\ref{eq:U_i}$), we can rewrite the utility function of TD $i$ as
\begin{align}
\label{eq:U_i_new}
U_i =& w_i ln(1+l_i)+v_i \\
&-(\gamma q_i \phi_i(L_i-l_i)+\gamma p_i\frac{l_i}{R_i^B})-d_i\phi_il_i.\notag
\end{align}
Then, the first-order derivative of $U_i$ can be calculated as
\begin{align}
\frac{\partial U_i}{\partial l_i} = \frac{w_i}{1+l_i} +\gamma q_i \phi_i - \gamma\frac{p_i}{R_i^B}-d_i\phi_i.
\label{eq:partial U_i}
\end{align}
Next, the second-order derivative of $U_i$ can be calculated as
\begin{align}
\frac{\partial^2U_i}{\partial l_i^2} = -\frac{w_i}{(1+l_i)^2}.
\label{eq: partial^2U_i}
\end{align}

It can be seen that the second-order derivative of $U_i$ is negative. Consequently, the utility function of TD $i$ exhibits strict concavity, indicating the existence of a unique optimal value of $l_i$ that maximizes $U_i$.

By setting the first-order derivative of $U_i$ to zero, we can derive the optimal offloading strategy for TD $i$ as
\begin{align}
l_i^{*} = \frac{w_iR_i^B}{\gamma p_i + d_i\phi_iR_i^B -\gamma q_i \phi_iR_i^B} -1.
\label{eq:l_i^{*}}
\end{align}

Subsequently, we will derive the optimal pricing strategy for the ES. Given the assumption of sufficient computational resources for the ES, the expression for $U_B$ undergoes transformation to $U_B = \sum_{i=1}^N U_{B,l}^i = \sum_{i=1}^N (d_i\phi_il_i - \gamma q_B\phi_il_i)$.

By substituting the optimal offloading strategy $\boldsymbol{l}^*$ for TDs obtained in the first stage into $U_B$, we can rewrite $U_B$ as
\vspace{-0.2cm}
\begin{align}
U_B  = \sum_{i=1}^N (d_i\phi_i-\gamma q_B\phi_i)(\frac{wR_i^B}{\gamma p_i + d_i\phi_iR_i^B -\gamma q_i \phi_iR_i^B} -1).
\label{eq:U_B_new}
\end{align}

Then, the first-order derivative of $U_B$ is formulated as
\begin{align}
\label{eq:partial U_B}
\frac{\partial U_B}{\partial d_i} =& \phi_i(\frac{w_iR_i^B}{\gamma p_i + d_i\phi_iR_i^B -\gamma q_i \phi_iR_i^B} -1)-\\
&(d_i\phi_i-\gamma q_B\phi_i) \frac{w_i\phi_i{R_i^B}^2}{(\gamma p_i + d_i\phi_iR_i^B -\gamma q_i \phi_iR_i^B)^2}.\notag
\end{align}

Next, the second-order derivative of $U_B$ is calculated as
\begin{align}
\frac{\partial^2 U_B}{\partial d_i^2} = \frac{-2w_i{\phi_i}^2 {R_i^B}^2[\gamma \phi_iR_i^B(q_B-q_i)+\gamma p_i]}{(\gamma p_i + d_iR_i^B -\gamma q_i \phi_iR_i^B)^3}.
\label{eq111}
\end{align}

Since $q_B-q_i>0$, as stated in Section II-C, this implies that the second-order derivative of $U_B$ is negative. Consequently, the utility function of the ES demonstrates strict concavity.

Similarly, the first-order derivative of $U_B$ is set to 0, 
\begin{align}
\frac{\partial U_B}{\partial d_i} = 0,
\label{eq: partial U_B = 0}
\end{align}
then we can derive the optimal pricing strategy of the ES for TD $i$. However, the equation is complex and nonlinear, making it challenging to find a closed-form optimal solution for $d_i$ directly. As $\frac{\partial^2U_B}{\partial d_i^2}<0$, indicating $\frac{\partial U_B}{\partial d_i}$ is strictly monotonic, we can employ the bisection search method to obtain a near-optimal solution for $d_i$.

At this point, we have obtained the optimal pricing strategy of the ES for each TD when the ES has sufficient computing resources. Next, we need to consider the constraint of the ES's computational capacity. When the computational resources of the ES are insufficient, certain tasks can be delegated to ADs for processing, with corresponding fees remunerated to the ADs. Moreover, typically, higher asking prices result in less data offloaded by TDs. Hence, the ES can raise the asking price to incentivize TDs to offload less data, ensuring that the idle resources of the ES and ADs meet the time constraints for all offloaded tasks.

As mentioned above, when the computational resources of the ES are insufficient, we need to solve two problems: determining task allocation among ADs and deciding whether and by what margin to adjust the asking price of the ES for certain TDs. To address these two issues, we will give the specific service pricing and task allocation strategy in the next section.

\section{Proposed Solution}\label{sec:Solution}
Recalling the problem analysis in the previous section, we first need to determine whether the computational resources of the ES are sufficient. Referring to Sections $\ref{sec:comp_model}$ and $\ref{sec:analysis}$, we can determine the  amount of computing resources that the ES needs to allocate to TD $i$ is $f_i^B(l_i^*) = \frac{\phi_il_i^*}{t_i^{max}-t_{i, B}^{tran}}$, assuming the ES has sufficient resources. Therefore, we can get the total computational resources that the ES needs to provide as $f_B =\sum_{i=1}^{N}f_i^B(l_i^*)$. If $f_B>F_B$, the ES's computational resources are insufficient. At this point, we need to assign some tasks to ADs for processing and raise the asking price of the ES for certain TDs to incentivize them to offload less data.

A priority-based strategy is employed to screen tasks designated for processing on ADs and determine their processing locations. The priorities of offloaded tasks and ADs are defined as
\vspace{-0.3cm}
\begin{align}
O_i = \frac{ d_i\phi_il_i^* - \gamma q_B\phi_il_i^*}{f_i^B(l_i^*)},
\label{eq:O_i}
\end{align}
and
\vspace{-0.4cm}
\begin{align}
Q_j^i = U_{B,j}^i(l_i^*),
\label{eq:Q_j^i}
\end{align}
respectively.
Here, $O_i$ represents the priority of task $h_i$ for processing on the ES. A higher value indicates higher priority, which corresponds to higher utility per unit of computational resource used by the ES to process this task. $Q_j^i$ signifies the priority of AD $j$ for handling task $h_i$. A higher value implies higher priority, which corresponds to higher utility when the ES delegates task $h_i$ to AD $j$ for processing.

In terms of increasing the asking price, it is crucial to determine the maximum threshold of the asking price for each task. Exceeding this threshold would dissuade users from offloading any data. Thus, by substituting $l_i^*=0$ into equation ($\ref{eq:l_i^{*}}$), we can calculate the maximum asking price for TD $i$ as
\begin{align}
d_i^{max} = \frac{w_i}{\phi_i}+\gamma(q_i-\frac{p_i}{\phi_iR_i^B}).
\label{eq:d_i^max}
\end{align}

Given the definitions and analysis provided above, Algorithm 1 is presented.

\IncMargin{0.5em}
\begin{algorithm}[h] \SetKwData{Left}{left}\SetKwData{This}{this}\SetKwData{Up}{up} \SetKwFunction{Union}{Union}\SetKwFunction{FindCompress}{FindCompress} \SetKwInOut{Input}{input}\SetKwInOut{Output}{output}

	\Input{$F_B$, $F_j$, $H_i$, $g_i^B$ and other parameters} 
	\Output{$\boldsymbol{d}^*$, $\boldsymbol{x}^*$, and the maximum offloading data volume constraint $\boldsymbol{l}^*$}

    Calculate $d_i$ and $d_{i}^{max}$ for each TD according to (\ref{eq: partial U_B = 0}) and (\ref{eq:d_i^max}), $\Delta_{d_i}= d_{i}^{max} - d_i $\;
    
    Calculate $ l_i $ and $ f_i^B(l_i) $ according to equations (\ref{eq:l_i^{*}}) and (\ref{eq:f_i^B}). $ f_B =\sum_{i=1}^{N}f_i^B(l_i)$\;
    \If{$f_B \le F_B$}{
        $d_i^*=d_i, x_i^*=0, l_i^*=l_i, \forall i\in \mathcal{N}$\;
    }\Else{
        Recruit ADs in the manner of Vickrey auction\;
        Initialize $f_B^{left}=F_B$, $f_j^{left}=F_j, \forall j\in \cal M$ \;
        Calculate $O_i, \forall i \in \cal N$ according to (\ref{eq:O_i}) and sort TDs in descending order to get $\mathcal{R}_{TD}=\{r_1,r_2,…,r_N\}$, where $r_i \in \mathcal{N}$ and $ O_{r_i}> O_{r_{i+1}} $\;
        \For{$q = 1,2, ... N$}{
            \If{$f_{r_q}^B(l_{r_q}) \le f_B^{left}$}{
                $d_{r_q}^*=d_{r_q}$, $x_{r_q}^*=0$, $l_{r_q}^*=l_{r_q}$, $ f_B^{left} = f_B^{left} - f_{r_q}^B$\;
            }\Else{
                Calculate $f_{r_q}^j(l_{r_q}) $ and $Q_j^{r_q}, \forall j \in \mathcal{M}$ according to (\ref{eq:f_i^j}) and (\ref{eq:Q_j^i}), and filter out the ADs that satisfy $f_j^{left} \ge f_{r_q}^j(l_{r_q})$ and $ Q_j^{r_q}\ge 0$ to form a set $\mathcal{M}^{\prime}$\;
                \If{$\mathcal{M}^{\prime} \neq \phi$}{
                    Find the AD  $j^*\in \mathcal{M}^{\prime}$ with the highest priority\;
                    $d_{r_q}^*=d_{r_q}$, $x_{r_q}^*=j^*$, $l_{r_q}^*=l_{r_q}$, $ f_{j^*}^{left} = f_{ j^*}^{left} - f_{r_q}^{ j^*}(l_{r_q})$\;
                    }\Else{
                        $d_{r_q}= d_{r_q} + \Delta_{d_{r_q}}/L$,  update $l_{r_q}$ and $ f^B_{r_q}(l_{r_q}) $ according to (\ref{eq:l_i^{*}}) and (\ref{eq:f_i^B})\;
                        \If{$ d_{r_q} \ge d_{r_q}^{max} $}{
                            $x_{r_q}^*=0, d_{r_q}^*=d_{r_q}^{max}, l_{r_q} ^*=0$\;
                            $\mathbf{continue}$\;
                        }
                        $\mathbf{goto}$ line 10\;
                    }
                
            }
        }
    }
    Return $\boldsymbol{d}^*$, $\boldsymbol{x}^*$, $\boldsymbol{l}^*$\;
	\caption{Proposed Algorithm Based on Stackelberg Model and Prioritization}\label{algo1}
    \end{algorithm} 
\DecMargin{0.5em} 
Algorithm 1 allocates task processing positions based on task and AD priorities. It encourages TDs to offload less data by increasing the asking price gradually. Additionally, Algorithm 1 imposes a constraint on the maximum data each TD can offload to prevent irrational users from causing shortages in supply-side resources through excessive data offloading.

\textit{Algorithm Complexity Analysis:} In Algorithm 1, we first solve $N$ equations using the bisection method. Assuming the accuracy requirement is $\epsilon$, the complexity is $\mathcal{O}(Nlog(\frac{1}{\epsilon}))$. Then, the TDs and ADs are sorted according to priorities with time complexities of $\mathcal{O}(NlogN)$ and $\mathcal{O}(MlogM)$, respectively. Additionally, the maximum number of price increments is $L$, which is a constant. Therefore, the overall time complexity of Algorithm 1 is $\mathcal{O}(NMlogNlogM)$.
% \usepackage{graphicx}

\begin{table}[htbp]
\centering
\caption{Main simulation parameters
}
\label{tab1}
\resizebox{\columnwidth}{!}{%
\begin{tabular}{l|ll}
\cline{1-2}
\textbf{Parameter}                       & \textbf{Value} &  \\ \cline{1-2}
Number of TDs $N$                        & [100, 160]     &  \\
Number of ADs $M$                        & [0, 30]        &  \\
Range of tolerant delay $t_i^{max}$      & [0.05 s, 3s]   &  \\
The size of computation tasks $L_i$      & [10 M, 20 M]   &  \\
The transmission power of TD $i$ $p_i$      & 0.1 W          &  \\
The fading component $\mu_0$             & 10             &  \\
Background noise $N_0$                   & -100 dBm       &  \\
Unit energy cost ($\$$/J)                   & 1              &  \\
Computation capacity of the ES $F_B$ & 10 GHz         &  \\
Computation capacity of AD $j$ $F_j$        & [1, 2] GHz     &  \\ \cline{1-2}
\end{tabular}%
}
\end{table}
\vspace{-0.3cm}
\section{Numerical Results}\label{sec:numerical}
This section presents simulation results to assess the performance of our proposed algorithm. The network scenario depicted in Fig. 1 is simulated. Table I lists the primary simulation parameters, some of which have been utilized in \cite{14}. All the above parameters will not change unless otherwise stated. Furthermore, we assume all TDs are rational entities, meaning they decide the quantity of offloaded data based on maximizing their utility. For comparison, we also simulate the following three intuitive strategies as benchmarks.

\begin{enumerate}
\item{\textbf{Uniform Pricing (UP):} The ES has consistent pricing for all TDs.}
\item{\textbf{Non-recruitment of ADs (NR):} The ES does not recruit ADs to participate in task processing.}
\item{\textbf{No Prioritization and Price Increment (NPPI):} Prioritization and price increment are not factored into the determination of pricing and task allocation strategies.}
\end{enumerate}

\begin{figure}[h]
\centering
\subfloat[]{\includegraphics[width=0.25\textwidth]{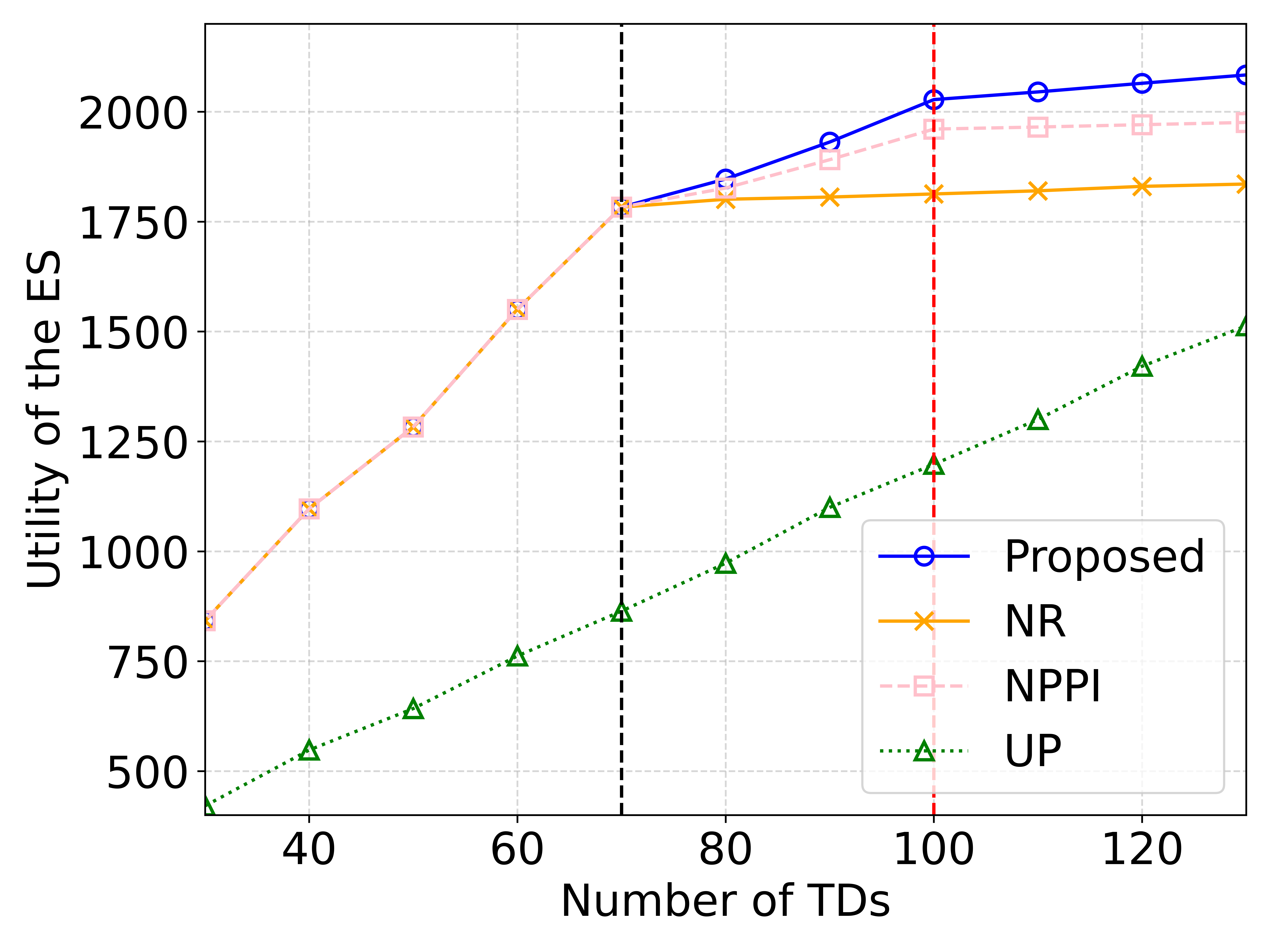}%
}
\subfloat[]{\includegraphics[width=0.25\textwidth]{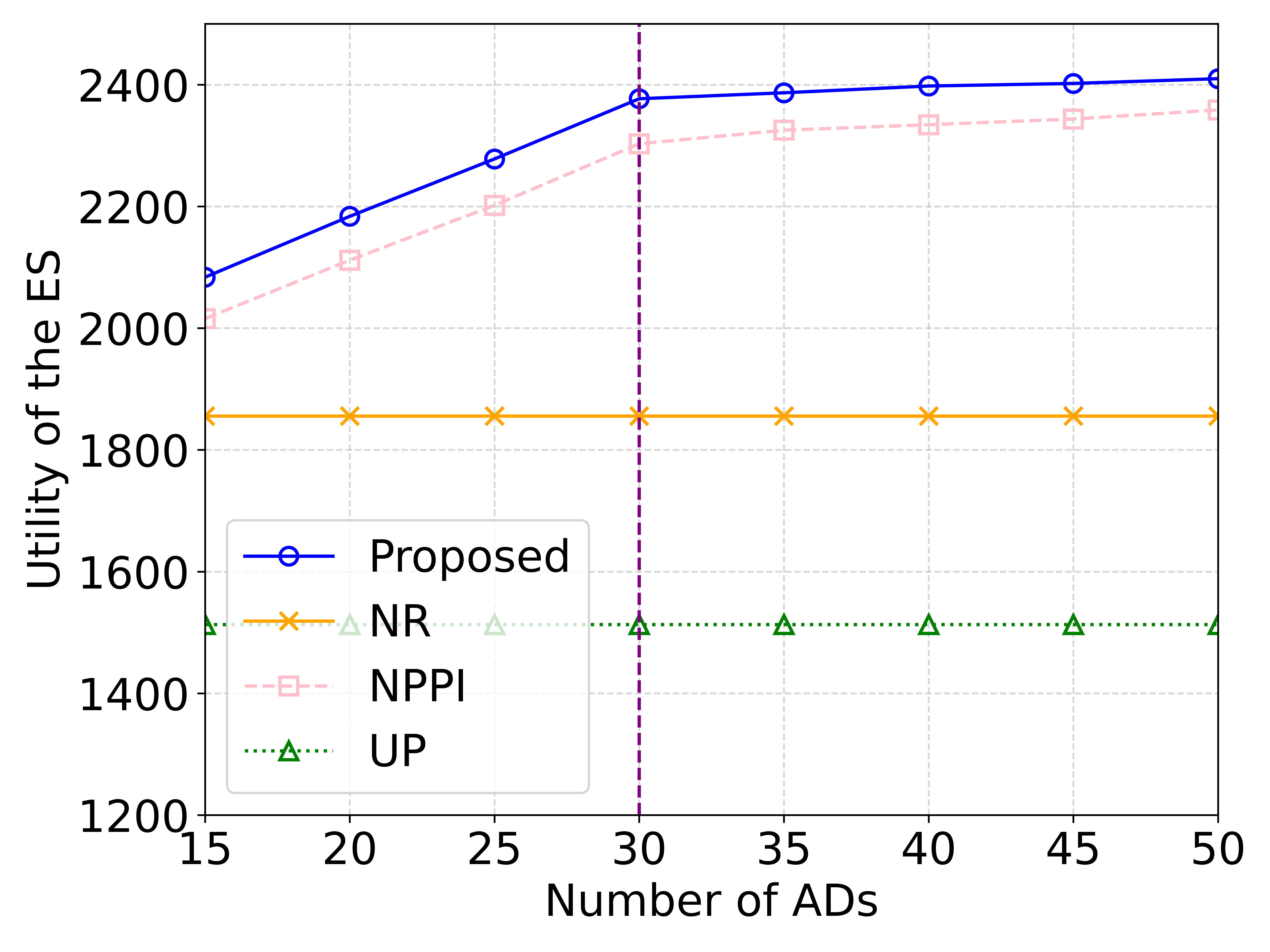}%
}
\vspace{-0.2cm}
\caption{(a) The utility of the ES versus the number of TDs. (b) The utility of the ES versus the number of ADs.}
\label{fig12}
\end{figure}

Fig. 3 depicts the utility of the ES under different strategies. In Fig. 3(a), the utility of the ES under the UP strategy consistently lags behind the other three strategies due to insufficient incentives for TDs. In Fig. 3(b), the utility of the ES under the NR and UP strategies remains constant as the number of ADs increases due to the underutilization of ADs' resources. These two figures demonstrate that the proposed scheme shows more pronounced performance enhancement compared to the other schemes, especially with a larger number of TDs. And the proposed scheme can improve the utility of the ES by about 45$\%$ compared to the UP strategy.

\begin{figure}[htbp]
\vspace{-0.3cm}
\centering
\subfloat[]{\includegraphics[width=0.25\textwidth]{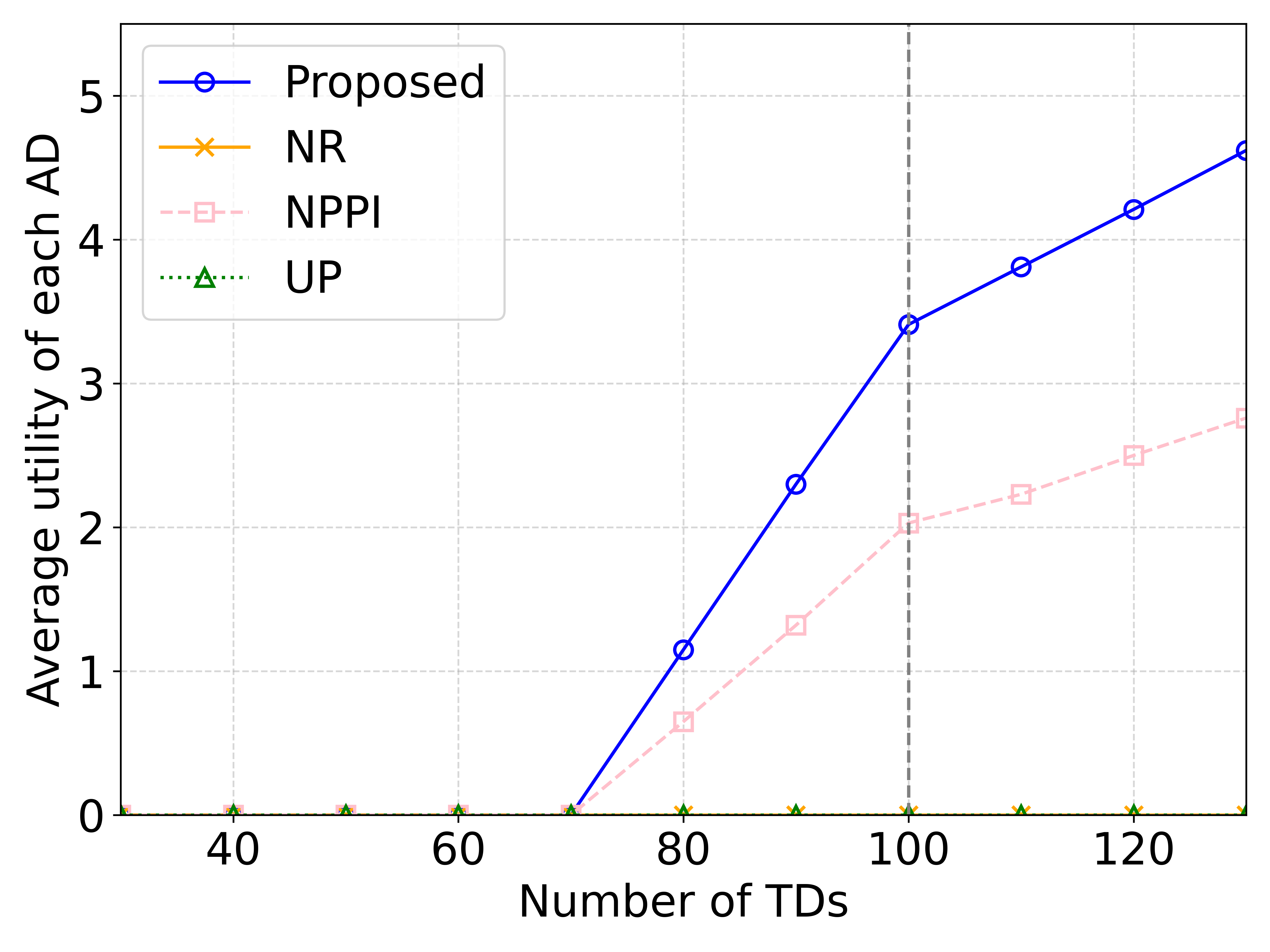}%
}
\subfloat[]{\includegraphics[width=0.25\textwidth]{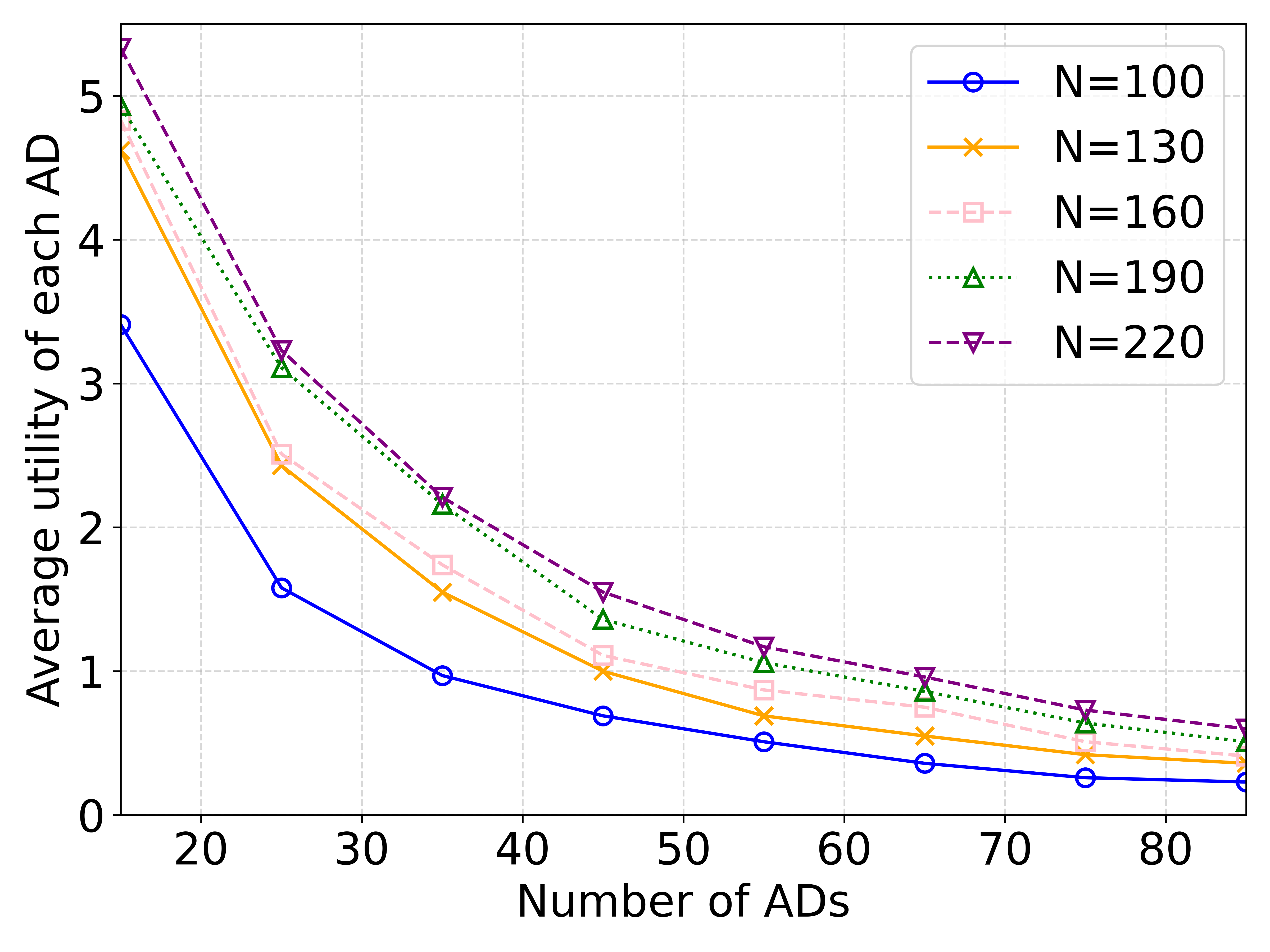}%
}
\vspace{-0.2cm}
\caption{(a) The average utility of each AD versus the number of TDs. (b) The average utility of each AD versus the number of ADs.}
\label{fig34}
\end{figure}
\vspace{-0.3cm}
Fig. 4 illustrates the average utility of each AD versus the number of TDs and ADs. As shown in Fig. 4(a), the average utility of each AD is greater than $0$ only when the resource demand of TDs surpasses the computational capacity of the ES. And the average utility of each AD under the proposed scheme is greater than or equal to the other schemes. Additionally, consistent with Fig. 3, both the proposed scheme and the NPPI strategy exhibit an inflection point at $N = 100$. Fig. 4(b) reveals that with inadequate resources of the ES, the average utility of each AD increases as the number of TDs rises and the number of ADs decreases. This phenomenon occurs because a higher number of TDs results in more resource demand on ADs, while a lower number of ADs reduces competition among them.

\begin{figure}[htbp]
\centering
\subfloat[]{\includegraphics[width=0.25\textwidth]{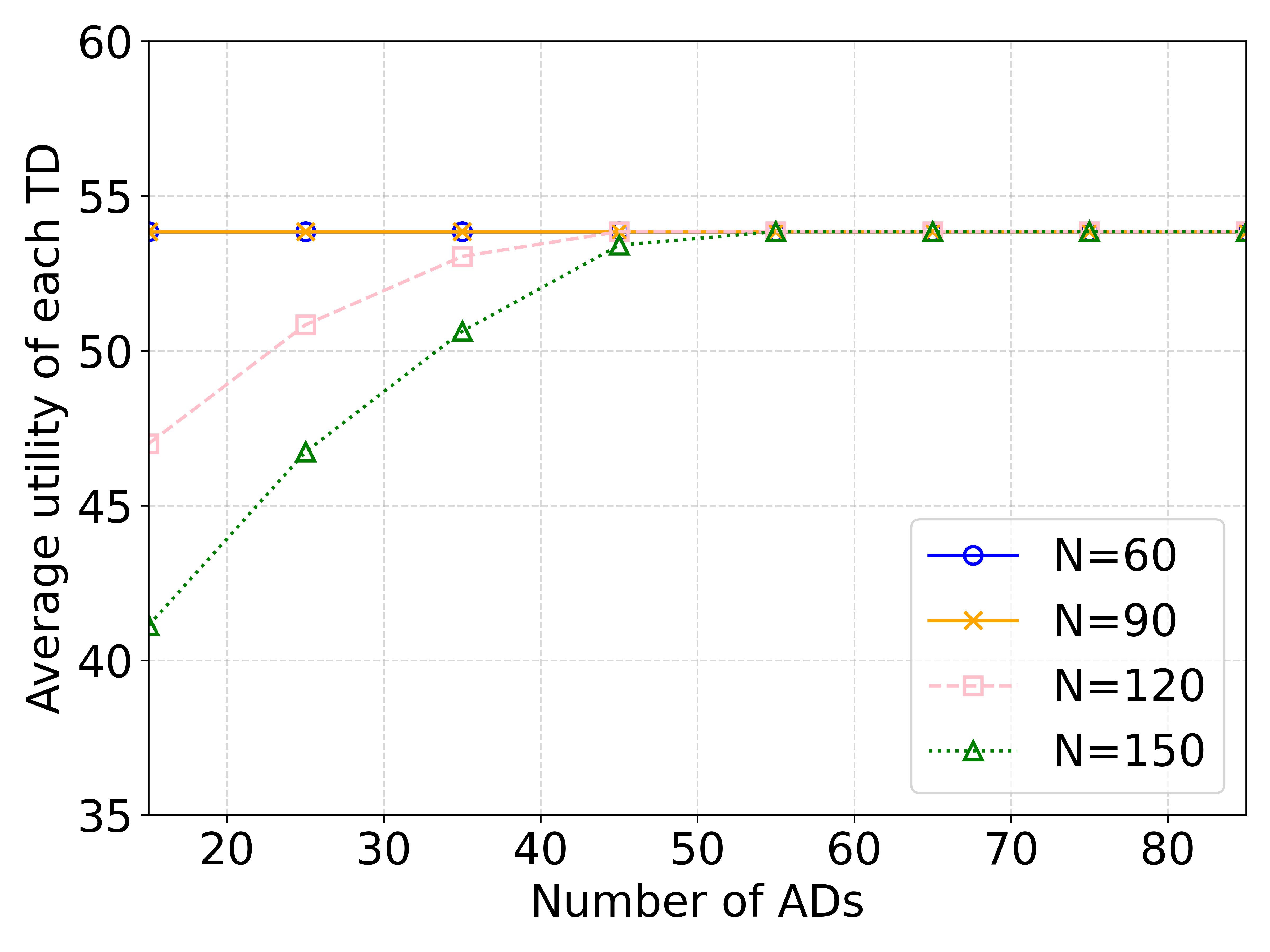}%
}
\subfloat[]{\includegraphics[width=0.25\textwidth]{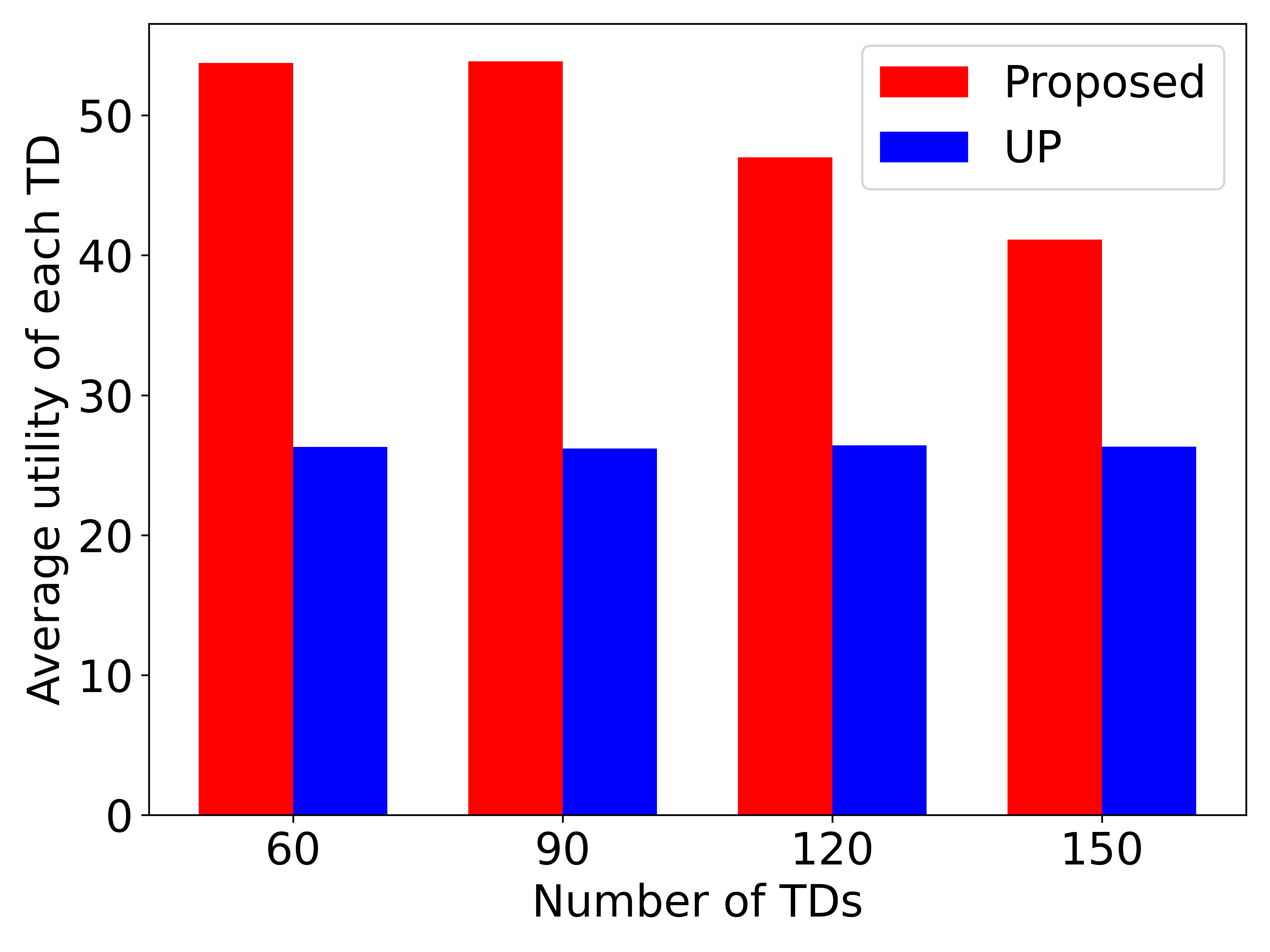}%
}
\vspace{-0.2cm}
\caption{(a) The average utility of each TD versus the number of ADs. (b) The average utility of each TD versus the number of TDs.}
\label{fig34}
\vspace{-0.3cm}
\end{figure}

Fig. 5 exhibits the average utility of each TD versus the number of ADs and TDs. From Fig. 6(a), it is evident that at $N=60$ and $90$, the resources on the supply side are adequate, and increasing ADs does not impact the average utility of each TD. However, at $N = 120$ and $150$, the supply-side resources become inadequate. This results in an improvement in the average utility of each TD as the number of ADs increases until the supply-side resources become sufficient again. After this point, the average utility of each TD stabilizes. Additionally, Fig. 6(b) illustrates that the UP strategy results in uneven incentives among TDs, thereby lowering the average utility of each TD. Tailored incentives for all TDs, employing the Stackelberg model, can effectively enhance their inclination towards offloading, thereby elevating the average utility of each TD by almost $50\%$.

\section{Conclusion}\label{sec:conclusion}
This paper investigates a device-assisted MEC network. To optimize the utility of the ES while protecting the interests of ADs and TDs, we employ the Vickrey auction to price the resources of ADs and the Stackelberg game to model the ES-TDs interactions. To foster efficient collaboration among TDs, ADs, and the ES, we propose a service pricing and task allocation algorithm based on the Stackelberg model and prioritization. Simulation results demonstrate that the proposed scheme maximizes the utility of the ES and enhances that of TDs and ADs. Future work is in progress, which involves extending from a single cell to multiple cells and exploring collaboration among multiple cells.

\section*{Acknowledgment}

This work is supported by the National Science Foundation of China under Grant 62071063 and by the BUPT Excellent Ph.D. Students Foundation under Grant CX20241066.

\bibliographystyle{IEEEtran}
\bibliography{Globecom_v2}

\begin{thebibliography}{1}
\providecommand{\url}[1]{#1}
\csname url@samestyle\endcsname
\providecommand{\newblock}{\relax}
\providecommand{\bibinfo}[2]{#2}
\providecommand{\BIBentrySTDinterwordspacing}{\spaceskip=0pt\relax}
\providecommand{\BIBentryALTinterwordstretchfactor}{4}
\providecommand{\BIBentryALTinterwordspacing}{\spaceskip=\fontdimen2\font plus
\BIBentryALTinterwordstretchfactor\fontdimen3\font minus
  \fontdimen4\font\relax}
\providecommand{\BIBforeignlanguage}[2]{{%
\expandafter\ifx\csname l@#1\endcsname\relax
\typeout{** WARNING: IEEEtran.bst: No hyphenation pattern has been}%
\typeout{** loaded for the language `#1'. Using the pattern for}%
\typeout{** the default language instead.}%
\else
\language=\csname l@#1\endcsname
\fi
#2}}
\providecommand{\BIBdecl}{\relax}
\BIBdecl

\bibitem{2}
T.~Fang, F.~Yuan, L.~Ao, and J.~Chen, ``{Joint task offloading, D2D pairing,
  and resource allocation in device-enhanced MEC: A potential game approach},''
  \emph{IEEE Internet Things J.}, vol.~9, no.~5, pp. 3226--3237, Jul. 2021.

\bibitem{3}
U.~Saleem, Y.~Liu, S.~Jangsher, X.~Tao, and Y.~Li, ``{Latency minimization for
  D2D-enabled partial computation offloading in mobile edge computing},''
  \emph{IEEE Trans. Veh. Technol.}, vol.~69, no.~4, pp. 4472--4486, 2020.

\bibitem{17}
Y.~Li, X.~Zhang, B.~Lei, Q.~Zhao, M.~Wei, Z.~Qu, and W.~Wang, ``{Computation
  rate maximization for wireless powered edge computing with multi-user
  cooperation},'' \emph{IEEE Open J. Commun. Soc.}, vol.~5, pp. 965--981, 2024.

\bibitem{16}
Y.~Li, X.~Ge, B.~Lei, X.~Zhang, and W.~Wang, ``{Joint task partitioning and
  parallel scheduling in device-assisted mobile edge networks},'' \emph{IEEE
  Internet Things J.}, Dec. 2023.

\bibitem{14}
H.~Zhou, Z.~Wang, G.~Min, and H.~Zhang, ``{UAV-aided computation offloading in
  mobile-edge computing networks: A Stackelberg game approach},'' \emph{IEEE
  Internet Things J.}, vol.~10, no.~8, pp. 6622--6633, 2022.

\bibitem{15}
L.~Ma, X.~Wang, X.~Wang, L.~Wang, Y.~Shi, and M.~Huang, ``{TCDA: Truthful
  combinatorial double auctions for mobile edge computing in industrial
  Internet of Things},'' \emph{IEEE Trans. Mob. Comput.}, vol.~21, no.~11, pp.
  4125--4138, 2021.

\bibitem{10}
B.~Lin, X.~Chen, X.~Chen, Y.~Ma, and N.~N. Xiong, ``{SGCS: An intelligent
  stackelberg game-based computation offloading and resource pricing scheme in
  blockchain-enabled MEC for IIoT},'' \emph{IEEE Internet Things J.}, early
  access, Feb. 28, 2024.

\bibitem{11}
M.~Chen, H.~Wang, D.~Han, and X.~Chu, ``{Signaling-based incentive mechanism
  for D2D computation offloading},'' \emph{IEEE Internet Things J.}, vol.~9,
  no.~6, pp. 4639--4649, Aug. 2021.

\bibitem{13}
F.~Zeng, Q.~Chen, L.~Meng, and J.~Wu, ``{Volunteer assisted collaborative
  offloading and resource allocation in vehicular edge computing},'' \emph{IEEE
  Trans. Intell. Transp. Syst.}, vol.~22, no.~6, pp. 3247--3257, Mar. 2020.

\end{thebibliography}

\end{document}